\documentclass[twocolumn,showpacs,aps,prl,superscriptaddress]{revtex4}
\usepackage{array}
\usepackage{graphicx}
\usepackage{units}
\usepackage{amsmath}
\usepackage{epsfig}

\input pubboard/babarsym

\def\totalLumi{230.2}  
\def\SigFitYield{489}
\def\SigFitYieldError{55}

\def\PWRValSyst{0.006}
\def\PWRValStat{0.018}
\def\PWRVal{0.143}

\def\FDsbabarValNorm{14}
\def\FDsbabarValSyst{7}
\def\FDsbabarValStat{17}
\def\FDsbabarVal{283}

\def\BRbabarValNorm{0.66}
\def\BRbabarValSyst{0.26}
\def\BRbabarValStat{0.83}
\def\BRbabarVal{6.74}

\def\BRpdgValNorm{1.29}
\def\BRpdgValSyst{0.20}
\def\BRpdgValStat{0.63}
\def\BRpdgVal{5.15}

\def\FDspdgValNorm{31}
\def\FDspdgValSyst{6}
\def\FDspdgValStat{15}
\def\FDspdgVal{248}

\def\MissEMin{0.38}
\def\NuPCorrection{-0.06}
\def\MuHelicity{0.90}
\def\DssMomMin{3.55}
\def\TotalEffi{8.13}
\def\SigFitYieldGOF{8.9}
\def\dsphipirecoeffi{9.90}
\def\dsPhiPiDataFitYield{2093}
\def\dsPhiPiDataFitYieldError{99}
\def\dsPhiPiDataGOF{25.0}

\def\brdsphipiPDG{\ensuremath{\unit[(3.6\pm0.9)]{\%}}}

\def\dstomunu{\ensuremath{\Ds\to\mu^+\nu_\mu}\xspace}
\def\dstophipi{\ensuremath{\Ds\to\phi\pi^+}\xspace}
\def\dsstodstophipi{\ensuremath{\Dss\to\gamma\Ds\to\gamma\phi\pi^+}\xspace}
\def\CSOne{\ensuremath{D^{*0}\to\gamma D^0\to\gamma\Km\pip}\xspace}

\newcommand{\BABARPubYear}    {06}
\newcommand{\BABARPubNumber}  {010}
\newcommand{\SLACPubNumber} {11968}

\long\def\inst#1{\par\nobreak\kern 4pt\nobreak
    {\it #1}\par\vskip 10pt plus 3pt minus 3pt}

\begin{document}

\preprint{\babar-PUB-\BABARPubYear/\BABARPubNumber} 
\preprint{SLAC-PUB-\SLACPubNumber} 

\begin{flushleft}
\babar-PUB-\BABARPubYear/\BABARPubNumber\\
SLAC-PUB-\SLACPubNumber\\
\end{flushleft}

\title{
  \vspace*{-1.0em}
  \large \bf \boldmath Measurement of the Pseudoscalar Decay Constant
  \fds Using Charm-Tagged Events in \epem Collisions at $\sqrt{s}=10.58\,\gev$ }

%
\author{B.~Aubert}
\author{R.~Barate}
\author{M.~Bona}
\author{D.~Boutigny}
\author{F.~Couderc}
\author{Y.~Karyotakis}
\author{J.~P.~Lees}
\author{V.~Poireau}
\author{V.~Tisserand}
\author{A.~Zghiche}
\affiliation{Laboratoire de Physique des Particules, F-74941 Annecy-le-Vieux, France }
% ...
\collaboration{The \babar\ Collaboration}
\noaffiliation

\date{\today}

\begin{abstract}
  \noindent
  Using \unit[\totalLumi]{\invfb} of \epem annihilation data collected
  with the \babar detector at and near the peak of the \Y4S resonance,
  $\SigFitYield\pm\SigFitYieldError$ events containing the pure
  leptonic decay $\Ds\to\mu^+\nu_\mu$ have been isolated in
  charm-tagged events. The ratio of partial widths
  $\Gamma(\dstomunu)/\Gamma(\dstophipi)$ is measured to be
  $\PWRVal\pm\PWRValStat\pm\PWRValSyst$ allowing a determination of
  the pseudoscalar decay constant $\fds =
  \unit[(\FDsbabarVal\pm\FDsbabarValStat\pm\FDsbabarValSyst\pm\FDsbabarValNorm)]{\mev}$.
  The errors are statistical, systematic, and from the \dstophipi
  branching ratio, respectively.
\end{abstract}


\maketitle

\noindent
Measurements of pure leptonic decays of charmed pseudoscalar
mesons are of particular theoretical importance. They provide an unambiguous
determination of the overlap of the wavefunctions of the heavy and light
quarks within the meson, represented by a single decay constant ($f_M$)
for each meson species ($M$). The partial width for a \Ds meson to
decay to a single lepton flavor ($l$) and its accompanying
neutrino ($\nu_l$), is given by
\begin{equation}
  \!\hspace{.3ex}\Gamma(\Ds\to l^+\nu_l)
  = \frac{G^2_F|V_{cs}|^2}{8\pi}  \fds^2m_l^2m^{}_{D_s}\Bigg(1-\frac{m^2_l}{m^2_{D_s}}\Bigg)^2,
  \label{eq:br}
\end{equation}
where $m^{}_{D_s}$ and $m^{}_l$ are the $\Ds$ and lepton masses,
respectively, $G_F$ is the Fermi constant, and $V_{cs}$ is the CKM
matrix element giving the coupling of the weak charged current to the $c$ and $s$ quarks
\cite{footnote:cc}.  The partial width is governed by two opposing
terms in $m^2_l$. The first term reflects helicity suppression in the
decay of the spin-0 meson, requiring the charged lepton to be in
its unfavored helicity state.  The second term is a phase-space
factor. As a result, the ratio of $\tau:\mu:e$ decays is approximately
$10:1:0.00002$. Lattice calculations have resulted in
$\fds=\unit[(249\pm17)]{\mev}$ and a ratio $\fds/\fsubd=1.24\pm0.07$ \cite{Aubin:2005ar}.
CLEO-c has recently measured a value for
$\fsubd=\unit[(223\pm17)]{\mev}$ \cite{Artuso:2005ym}.

We present herein the most precise measurement to date of the ratio
$\Gamma(\dstomunu)/\Gamma(\dstophipi)$ and the decay
constant \fds. The data (\unit[\totalLumi]{\invfb}) were collected
with the \babar detector at the asymmetric-energy $\epem$ storage ring
\pep2 at and below the \Y4S resonance. The \babar detector is
described in detail elsewhere \cite{babar:2002}. Briefly, the
components used in this analysis are the tracking system composed of a
five-layer silicon vertex detector and a 40-layer drift chamber (DCH),
the Cherenkov detector (DIRC) for charged $\pi$--$K$
discrimination, the CsI(Tl) calorimeter (EMC) for photon and electron
identification, and the 18-layer flux return (IFR) located outside the
\unit[1.5]{T} solenoid coil and instrumented with resistive plate
chambers for muon identification and hadron rejection.

The analysis proceeds as follows. In order to measure
$\Ds\to\mu^+\nu_\mu$, the decay chain $\Dss\to\gamma\Ds,
\Ds\to\mu^+\nu_\mu$ is reconstructed from \Dss mesons produced in the
hard fragmentation of continuum \ccbar events.  The subsequent decay
results in a photon, a high-momentum \Ds and daughter muon and
neutrino, lying mostly in the same hemisphere of the event.  Signal
candidates are required to lie in the recoil of a fully reconstructed
\Dz, \Dp, \Ds, or \Dstarp meson (the ``tag''), wherein the tag flavor, and
hence the expected charge of the signal muon, is uniquely determined.
To eliminate signal from \B decays, the minimum tag momentum is chosen
to be close to the kinematic limit for charm mesons arising from \B
decays. Tagging in this manner significantly reduces backgrounds,
while improving the missing mass resolution of the signal.

\begin{figure}[tbp]
  \centering
  \epsfig{file = 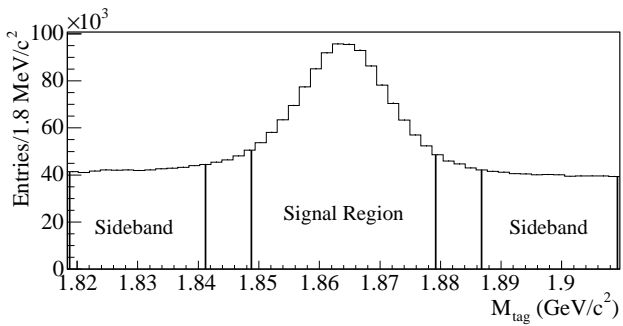, width=\columnwidth}
  \caption{
    Tag mass distribution, showing the signal and sideband regions, in
    events with a recoil muon.  All tag modes are combined, scaling
    their mass and width to that of the $\Dz\to\Km\pip$ mode.  }
  \label{fig:tagmass}
\end{figure}

Tag candidates are reconstructed in the following modes:
$\Dz\to\Km\pip(\piz),\Km\pip\pip\pim$,
$\Dp\to\Km\pip\pip(\piz),\KS\pip(\piz),\KS\pip\pip\pim,\Kp\Km\pip,\KS\Kp$,
$\Ds\to\KS\Kp,\phi\rho^+$, and $\Dstarp$ $\to$ $\pip\Dz$, with $\Dz$
$\to$ $\KS\pip\pim(\piz), \KS\Kp\Km,\KS\piz$. Kaons are identified
using information from the DCH and the DIRC.  Requirements on the
vertex probability of the tag decay products are imposed. For each tag
mode a signal region and sideband regions in the tag mass distribution
are defined.  The signal region spans $\pm2$ standard deviations
($\sigma_{\rm tag}$) around the mean ($\mu_{\rm tag}$), determined
from fits to the tag mass distribution in data events.  The sidebands
extend from 3 to 6 $\sigma_{\rm tag}$ on either side of $\mu_{\rm
  tag}$ (Fig.~\ref{fig:tagmass}).

For each event a single tag candidate is chosen and then used in the
subsequent analysis.  To pick this tag among multiple candidates
within an event (there are 1.2 candidates on average in events with at
least one candidate) modes of higher purity are preferred. In events
where two tag candidates are reconstructed in the same mode, the
quality of the vertex fit of the $D$ meson is used as a secondary
criterion. After subtracting combinatorial background there are
$5*10^5$ charm tagged events with a muon amongst the recoiling
particles.

The signature of the decay $\Dss\to\gamma\Ds$ is a narrow peak in the
distribution of the mass difference $\Delta M =
M(\mu\nu\gamma)-M(\mu\nu)$ at \unit[143.5]{\mevcc}. The \Dss signal is
reconstructed from a muon and a photon candidate in the recoil of the
tag. Muons are identified as non-showering tracks penetrating the IFR.
The muon must have a momentum of at least \unit[1.2]{\gevc} in the
center-of-mass (CM) frame and have a charge consistent with the tag
flavor. Muons used in this analysis are identified with an average
efficiency of $\unit[\approx70]{\%}$, while the pion misidentification
rate is $\unit[\approx2.5]{\%}$. Clusters of energy in the EMC not
associated with charged tracks are identified as photon candidates.
The photon CM energy must exceed $\unit[0.115]{\gev}$.

The CM missing energy ($E^*_\text{miss}$) and momentum ($\vec
p^*_\text{miss}$) are calculated from the four-momenta of the incoming
\epem, the tag four-momentum, and the four-momenta of all remaining
tracks and photons in the event. The energy of the charged particles
that do not belong to the tag is calculated from the track momentum
under a pion mass hypothesis. Assigning a mass according to the most
likely particle hypothesis has negligible effect on the missing energy
resolution. Since the neutrino in the signal decay leads to a large
missing energy in the event, the requirement
$E^*_\text{miss}>\unit[\MissEMin]{\gev}$ is made.

The neutrino CM four-momentum ($p^*_\nu=(|\vec p^*_\nu|,\vec
p^*_\nu)$) is estimated from the muon CM four-momentum ($p^*_\mu$) and
$\vec p^*_\text{miss}$, using a technique adopted from
Ref.~\cite{Chadha:1998zh}. The difference $|\vec p^*_{\rm miss} - \vec
p^*_\nu|$ is minimized, while the invariant mass of the neutrino-muon
pair is required to be the known mass of the \Ds
\cite{Eidelman:2004wy}. Studies of simulated decays of signal and
background \ccbar events show that the quantity $p_\text{corr} = |\vec
p^*_{\rm miss}| - |\vec p^*_\nu|$ is centered at 0 for signal decays,
while for the \ccbar events it peaks at a negative value significantly
separated from the signal. A requirement
$p_\text{corr}>\NuPCorrection\,\gevc$ is imposed. To reduce
contributions from background events where particles are lost along
the beam pipe in the forward direction, a requirement on the neutrino
CM polar angle $\theta^*_\nu>\unit[38]{\degrees}$ is made. The muon CM
four-momentum ($p^*_\mu$) is combined with $p^*_\nu$ to form the \Ds
candidate.  Unlike the signal \Ds, a large number of random \Ds
combinations have the muon candidate aligned with the \Ds flight
direction. A requirement $\cos(\alpha_{\mu,D_s})<\MuHelicity$ is made
on the angle between the muon direction in the \Ds frame and the \Ds
flight direction in the CM frame. The \Ds candidate is then combined
with a photon candidate to form the \Dss.  The CM momentum of
correctly reconstructed \Dss is typically higher than that of random
combinations; signal candidates are required to have $|\vec
p^*_{\Dss}|>\DssMomMin\,\gevc$.  The resulting signal detection
efficiency in tagged events is
$\epsilon_\text{Sig}=\unit[\TotalEffi]{\%}$.

The selection requirements on $E^*_\text{miss}$, $\alpha_{\mu,D_s}$,
$p_\text{corr}$, $\theta_\nu^*$, and $|\vec p^*_{\Dss}|$ are optimized
using simulation to maximize the significance $s/\sqrt{s+b}$, where
$s$ and $b$ are the signal and background yields expected in the data
set. Backgrounds arise from several distinct sources.  The first
class of background are events $\epem\to f\bar f$, where $f =
u,d,s,b$, or $\tau$, which do not contain a real charm tag. The
contribution of these events is estimated from data using the tag
sidebands. In addition there are events $\epem\to\ccbar$ where the tag
is incorrectly reconstructed. Although these events potentially
contain the signal decay, they are also subtracted using the tag
sidebands.  These two sources amount to $\unit[\approx42]{\%}$ of the
background.

The second class of background events ($\unit[\approx26]{\%}$) are
correctly tagged $\ccbar$ events with the recoil muon coming from a
semileptonic charm decay or from $\tau^+\to\mu^+\nu_\mu\bar\nu_\tau$.
This includes events $\Dss\to\gamma\Ds\to\gamma\tau^+\nu_\tau,\ 
\tau^+\to\mu^+\nu_\mu\bar\nu_\tau$. To estimate the size and shape of
this background contribution, the analysis is repeated, substituting a
well-identified electron for the muon. Except for a small phase-space
correction, the widths of weak charm decays into muons and electrons
are assumed to be equal. QED effects such as bremsstrahlung
($e^+\to\gamma e^+$) energy losses and photon conversion ($\gamma\to
e^+e^-$), where the muon equivalents have a much lower rate, are
explicitly removed.  In particular, bremsstrahlung photons found in
the vicinity of an electron track are combined with the track.  The
small number of events with an electron from a converted photon that
survive the selection are suppressed by a photon conversion veto,
using the vertex and the known radial distribution of the material in the
detector.
The muon selection efficiency as a function of momentum and direction is
measured using $\epem\to\mu^+\mu^-\gamma$ events, while radiative
Bhabha events are used to quantify the electron efficiency. The ratio
of muon to electron efficiencies is applied as a weight to each
electron event.

The remaining backgrounds are estimated from simulation.  These
include events ($\unit[\approx20]{\%}$) with pure leptonic decays of a
\Ds or \Dp meson, $\Dp_{(s)}\to\mu^+\nu_\mu$, where the $\Dp_{(s)}$ is
produced either directly in \ccbar fragmentation or in decays of
$\Dstarp_{(s)}$, excluding the signal decay chain. If the photon used
in the reconstruction originates from a \piz of a $\Dstarp_{(s)}$
decay, the $\Delta M$ distribution peaks sharply around $70\,\mevcc$;
otherwise it is flat.  A small background ($\unit[\approx1]{\%}$)
arises from decays $\Dss\to\gamma\Ds\to\gamma\tau^+\nu_\tau$ with
$\tau^+\to\pip(\piz)\nu_\tau$ and the charged pion being misidentified
as a muon. Its $\Delta M$ distribution peaks close to that of the
signal.  Other backgrounds ($\unit[\approx10]{\%}$) include signal
events with an incorrectly chosen photon candidate, and hadronic
\ccbar events with one of the final state hadrons, usually a \pip or a
\Kp, being misidentified as a muon. These backgrounds have a flat
$\Delta M$ distribution.

\begin{figure}[tbp]
  \centering
  \epsfig{file=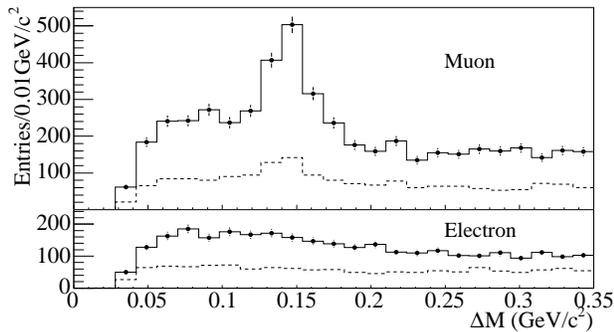,width=\columnwidth}
  \caption{$\Delta M$ distribution of charm-tagged events passing the signal
    selection.  The tag can be from the tag signal region (solid
    lines) or the sidebands (dashed lines). In the bottom plot the
    signal muon is replaced with an electron to estimate the
    semileptonic charm and $\tau$ decay background.
 }
  \label{fig:datafour}
\end{figure}

\begin{figure}[t]
  \centering
  \epsfig{file=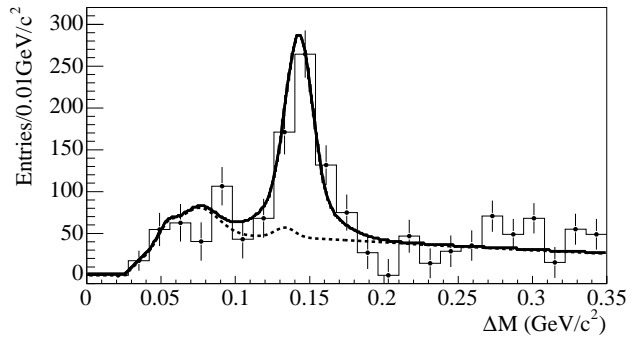,width=\columnwidth}
  \caption{$\Delta M$ distribution after the tag sidebands and
    the electron sample are subtracted. The solid line is the fitted
    signal and background distribution $(N_\text{Sig}f_\text{Sig} +
    N_\text{Bkgd}f_\text{Bkgd})$, the dashed line is the background
    distribution $(N_\text{Bkgd}f_\text{Bkgd})$ alone. }
  \label{fig:data}
\end{figure}

Events that pass the signal selection are grouped into four sets,
depending on whether the tag lies in the signal region or the sideband
regions, and on whether the lepton is a muon or an electron
(Fig.~\ref{fig:datafour}).  For each lepton type the sideband $\Delta
M$ distribution is subtracted.  The electron distribution, scaled by
the relative phase-space factor (0.97) appropriate to semileptonic
charm meson decays and leptonic $\tau$ decays is then subtracted from
the muon distribution.  The resulting $\Delta M$ distribution is
fitted with a function $(N_\text{Sig}f_\text{Sig} +
N_\text{Bkgd}f_\text{Bkgd})(\Delta M)$, where $f_\text{Sig}$ and
$f_\text{Bkgd}$ describe the simulated signal and background $\Delta
M$ distributions. The function $f_\text{Sig}$ is a double Gaussian
distribution. The function $f_\text{Bkgd}$ consists of a double and a
single Gaussian distribution describing the two peaking background
components, and a function\,\cite{argusfnc} describing the flat
background component. The relative sizes of the background components,
along with all parameters except $N_\text{Sig}$ and $N_\text{Bkgd}$
are fixed to the values estimated from simulation. The $\chi^2$ fit
yields $N_\text{Sig} = \SigFitYield\pm\SigFitYieldError(stat)$ signal
events and has a fit probability of \unit[\SigFitYieldGOF]{\%}
(Fig.~\ref{fig:data}).

The branching fraction of \dstomunu cannot be determined directly,
since the production rate of $D_s^{(*)+}$ mesons in \ccbar
fragmentation is unknown.  Instead the partial width ratio
$\Gamma(\dstomunu)/\Gamma(\dstophipi)$ is measured by reconstructing
\dsstodstophipi decays. The \dstomunu branching fraction is evaluated
using the measured branching fraction for \dstophipi.

\begin{figure}[t]
  \centering
  \epsfig{file=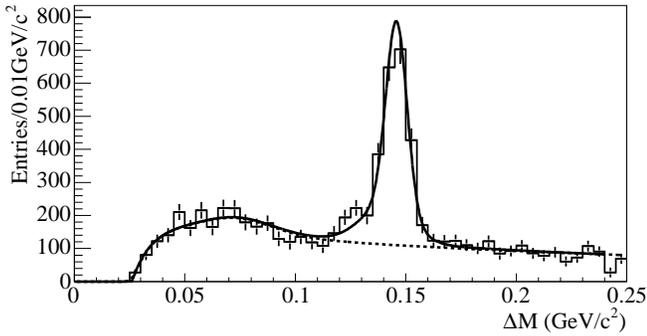,width=\columnwidth}
  \caption{$\Delta M$ distribution of selected \dsstodstophipi events
    after the tag sideband is subtracted. The solid line is the fitted
    signal and background distribution
    $(N_{\phi\pi}f_{\phi\pi} + N_{\phi\pi\text{Bkgd}}f_{\phi\pi \text{Bkgd}})$, the dashed line is the
    background distribution $(N_{\phi\pi \text{Bkgd}} f_{\phi\pi \text{Bkgd}})$ alone. }
  \label{fig:dsphipi}
\end{figure}

Candidate $\phi$ mesons are reconstructed from two kaons of opposite
charge. The $\phi$ candidates are combined with charged pions to form
\Ds meson candidates. Both times a geometrically constrained fit is
employed, and a minimum requirement on the fit quality is made. The
$\phi$ and the \Ds candidate masses must lie within $2\,\sigma$ of
their nominal values, obtained from fits to simulated events and data.
Photon candidates are then combined with the \Ds to form \Dss
candidates. The same requirements on the CM photon energy and \Dss
momentum as in the \dstomunu signal selection are made. The
\dsstodstophipi selection efficiency in tagged events is
$\epsilon_{\phi\pi}=\unit[\dsphipirecoeffi]{\%}$.  Data events that
pass the selection are grouped into two sets: the tag signal and
sideband regions. After the tag sideband has been subtracted from the
tag signal $\Delta M$ distribution, the remaining distribution is
fitted with $(N_{\phi\pi}f_{\phi\pi}+N_{\phi\pi \text{Bkgd}}f_{\phi\pi
  \text{Bkgd}})(\Delta M)$, where $f_{\phi\pi}$ is a triple Gaussian,
describing the simulated \dsstodstophipi signal, and $f_{\phi\pi
  \text{Bkgd}}$ consists of a broad Gaussian centered at $70\,\mevcc$
and a function\,\cite{argusfnc} describing the simulated background
$\Delta M$ distributions. The Gaussian describes the background
$\Dss\to\piz\Ds\to\piz\phi\pip$ where the photon candidate originates
from the \piz. The relative sizes of the background components, along
with all parameters except $N_{\phi\pi}$, $N_{\phi\pi\text{Bkgd}}$,
and the mean of the peak are fixed to the values estimated from
simulation.  The $\chi^2$ fit yields
$N_{\phi\pi}=\dsPhiPiDataFitYield\pm\dsPhiPiDataFitYieldError$ events
and has a probability of \unit[\dsPhiPiDataGOF]{\%}
(Fig.~\ref{fig:dsphipi}). From simulation $48\pm23$ events
$\Dss\to\gamma\Ds\to\gamma f_0(980)(\Kp\Km)\pip$ are expected to
contribute to the signal, where the error is mostly from the
uncertainty in the $\Ds\to f_0(980)(\Kp\Km)\pip$ braching ratio.

Precise knowledge of the efficiency of reconstructing the tag is not
important, since it mostly cancels in the calculation of the partial
width ratio. However, the presence of two charged kaons in \dstophipi
events leads to an increased number of random tag candidates, compared
to \dstomunu events, which decreases the chances that the correct tag
is picked. The size of the correction for this effect to the
efficiency ratio ($\epsilon_{\phi\pi}/\epsilon_\text{Sig}$) is
determined to be \unit[$-1.4$]{\%} in simulated events.

To measure the effect of a difference between the \Dss momentum
spectrum in simulated and data events, \dsstodstophipi events are
selected in data with the \Dss momentum requirement removed. The
sample is purified by requiring the CM momentum of the charged pion to
be at least \unit[0.8]{\gevc}. The efficiency-corrected \Dss momentum
distribution in data is compared to that of \Dss in simulated
\dsstodstophipi events. A harder momentum spectrum is observed in
data.  The detection efficiencies for signal and \dsstodstophipi
events are re-evaluated after weighting simulated events to match the
\Dss momentum distribution measured in data. The correction to the
efficiency ratio is \unit[$+1.5$]{\%}.
 
With both corrections applied, the partial width ratio is determined
to be $\Gamma_{\mu\nu}/\Gamma_{\phi\pi} = (N/\epsilon)_\text{Sig}/(N/\epsilon)_{\phi\pi}\times\BR(\phi\to K^+K^-) = \PWRVal \pm
\PWRValStat(stat)$, with $\BR(\phi\to K^+K^-)=\unit[49.1]{\%}$ \cite{Eidelman:2004wy}.

The combined systematic uncertainty due to the corrections applied,
taken as half the size of each correction, is \unit[1.0]{\%}.
The systematic error in the signal efficiency due to selection
criteria insensitive to the \Dss momentum is evaluated using
reconstructed \CSOne events. The conditions present in the signal are
emulated by removing the charged pion, taken to represent the
neutrino, from these events.  The signal reconstruction and selection
steps are repeated, and the selection efficiencies compared between
simulated and data events.  The assigned systematic uncertainty is
\unit[1.4]{\%}.  For the \dstophipi selection, requirements on the \Ds
and $\phi$ vertex fit probability contribute a systematic uncertainty
of \unit[0.7]{\%}, estimated from comparisons of \dstophipi events in
simulation and data. Control samples of $\epem\to\mu^+\mu^-\gamma$ and
$\Dstarp\to\pip\Dz\to\pip\Km\pip$ events are used to measure the
particle identification efficiencies of muons and charged kaons and
pions in data, and to correct the simulated signal and \dsstodstophipi
efficiencies. An uncertainty of \unit[0.7]{\%} is associated with
these corrections, mainly due to the limited statistics of the control
samples. The systematic uncertainties in the track reconstruction
efficiency cancel partially in the \dstomunu to \dstophipi ratio and
contribute \unit[1.2]{\%}.  An additional uncertainty of
\unit[1.1]{\%} is due to the statistical limitations of the simulated
signal and \dstophipi event samples.  Simulation studies are used to
evaluate the systematic uncertainties arising from a possible
inadequate parameterization of the signal (\unit[0.9]{\%}) and
background (\unit[2.3]{\%}) shapes. Simulations are also used to
determine the systematic uncertainty associated with the subtraction
of the electron sample (\unit[0.4]{\%}). The error on the branching
ratio $\BR(\phi\to K^+K^-)$ is 1.2\,\%, the uncertainty on the $\Ds\to
f_0(980)\pip$ background is 1.1\,\%.  The total systematic uncertainty
on $\Gamma(\dstomunu)/\Gamma(\dstophipi)$ is \unit[3.9]{\%}.

Using the \babar average for the branching ratio
$\BR(\dstophipi)=(4.71\pm0.46)\,\%\,$\cite{Aubert:2005xu}\cite{Aubert:2006nm},
we obtain the branching fraction $\BR(\dstomunu) = (\BRbabarVal \pm
\BRbabarValStat \pm \BRbabarValSyst \pm \BRbabarValNorm)\times10^{-3}$
and the decay constant $\fds = \unit[(\FDsbabarVal \pm
\FDsbabarValStat \pm \FDsbabarValSyst \pm \FDsbabarValNorm)]{\mev}$.
The first and second errors are statistical and systematic,
respectively; the third is the uncertainty from $\BR(\dstophipi)$. The
ratio of our value for \fds to \fsubd\ from the CLEO-c measurement,
$\fds/\fsubd=1.27\pm0.14$, is
consistent with lattice QCD.\\[0ex]
\hspace*{\parindent}Using $\BR(\dstophipi)_\text{PDG} = \brdsphipiPDG$
\cite{Eidelman:2004wy}, the branching fraction is $\BR(\dstomunu) =
(\BRpdgVal \pm \BRpdgValStat \pm \BRpdgValSyst \pm
\BRpdgValNorm)\times10^{-3}$ and the decay constant $\fds =
\unit[(\FDspdgVal \pm \FDspdgValStat \pm \FDspdgValSyst \pm
\FDspdgValNorm)]{\mev}$.

We are grateful for the excellent luminosity and machine conditions
provided by our \pep2\ colleagues, 
and for the substantial dedicated effort from
the computing organizations that support \babar.
The collaborating institutions wish to thank 
SLAC for its support and kind hospitality. 
This work is supported by
DOE
and NSF (USA),
NSERC (Canada),
IHEP (China),
CEA and
CNRS-IN2P3
(France),
BMBF and DFG
(Germany),
INFN (Italy),
FOM (The Netherlands),
NFR (Norway),
MIST (Russia), and
PPARC (United Kingdom). 
Individuals have received support from the
Marie Curie EIF (European Union) and
the A.~P.~Sloan Foundation.


\begin{thebibliography}{99}
\bibitem{footnote:cc} Charge conjugation is implied throughout this
  Letter.
\end{thebibliography}
\end{document}